# AN EMPIRICAL STUDY ON THE PROCEDURE TO DERIVE SOFTWARE QUALITY ESTIMATION MODELS

[1]Jie Xu, [2]Danny Ho and [1]Luiz Fernando Capretz

[1]Department of Electrical and Computer Engineering, University of Western Ontario, London, Ontario, N6A 5B9 Canada

jxu89@uwo.ca; lcapretz@eng.uwo.ca

[2]NFA Estimation Inc., Richmond Hill, Ontario, Canada

danny@nfa-estimation.com

## ABSTRACT

*Software quality assurance has been a heated topic for several decades. If factors that influence software quality can be identified, they may provide more insight for better software development management. More precise quality assurance can be achieved by employing resources according to accurate quality estimation at the early stages of a project. In this paper, a general procedure is proposed to derive software quality estimation models and various techniques are presented to accomplish the tasks in respective steps. Several statistical techniques together with machine learning method are utilized to verify the effectiveness of software metrics. Moreover, a neuro-fuzzy approach is adopted to improve the accuracy of the estimation model. This procedure is carried out based on data from the ISBSG repository to present its empirical value.*

## 1. INTRODUCTION

Software quality assurance is one of the most important components in software project management. Research on various perspectives of software quality and related activities has been conducted for several decades, and many conclusions and practices have been presented to improve software quality. One aspect of the research in this area is to establish software quality estimation models that could be used at the early stages of a project to estimate the quality level. The estimation results can act as a guideline to enhance the quality assurance performance.

Since the meaning of software quality has been interpreted differently, only some characteristics of software quality have been discussed to elaborate its definition. Like in ISO-9126, software quality is broken down into subclasses such as: functionality, reliability, usability, efficiency, maintainability and portability, which can be further measured by respective sub-characteristics.

Although hierarchical quality models present overall views of software quality, they are not popular for quantifiable management. Instead, direct measures of software quality are more preferable in current practical quality insurance. Among them, the number of defects found in a project is still a common indicator of software quality.

Although great efforts have been put on related research and some accomplishments have been made, there are still many questions unsolved about software quality assurance, especially software quality estimation models. After reviewing and analyzing the published papers and books in this area, we determine that the following research questions are crucial:

(1) How effective software metrics can be validated in managing software quality control and estimation?

(2) What techniques and approaches can be utilized to build software quality estimation models?

(3) Is there a general procedure one can follow to derive software quality estimation models?

In our study, we define a general procedure for deriving software quality estimation models, and suggest feasible techniques that could be adopted to accomplish the specific tasks in the respective steps of the procedure. During this process, the above research questions are addressed.

## 2. RELATED WORKS

To derive software quality estimation models for the number of defects, many researchers have proposed techniques and approaches to accomplish the goal, and various software metrics have been identified in those models. They performed estimation tasks at two levels: modules and projects. Many techniques were applied, such as linear regression [1], Case-Based Reasoning (CBR) [2], fuzzy logic [3], neural networks [4], Bayesian networks (BN) [5], etc.

Chidamber and Kemerer (CK) [6] introduced their OO design and complexity metrics and demonstrated the clear impact on software quality. Other variants of CK metrics were designed in order to present more accurate implications of software quality. Although all these studies made valuable contributions to improve OO design, their results were not consistent [7,8,9]. Other regression methods such as Poisson regression and zero-inflated Poisson were also adopted to build estimation models with complexity metrics [10]. On the other hand, other software metrics, such as Halstead software science [11], McCabe's cyclomatic complexity [12], were also designed to reveal their influence on software quality.

Fenton and Neil [13] reviewed a wide range of defect prediction models, which mostly relied on size and complexity metrics. They argued that those models lacked validation and had many theoretical and practical problems. They stated that Bayesian method could be a good option to solve the problem, but failed to provide solid proofs.

Chulani and Boehm [14] proposed a COnstructive QUALity MOdel(COQUALMO), which consisted of two sub-models: the Defect Introduction and the Defect Removal sub-models. The first sub-model used 21 defect introduction drivers together with the size of the project to derive total numbers of defects, and the second estimated residual defects with defect removal profiles. The techniques applied to build the model were regression and Delphi-gathered expert opinions. This model was an extension to COnstructive COst Model (COCOMO), but was not accepted as widely as COCOMO.

Bibi et al. [15] performed their research on estimating the number of defects through regression via classification. In their model, 5 classification variables, 10 quantitative variables and 10 risk factors were adopted as predictors. The model was compared with those derived from other machine learning techniques, but why the variables were selected was not explained.

These studies focused on solve specific estimation problems with particular techniques, but none of them explicitly described the modeling procedures that they have used to achieve their goals.

## 3. MODELING PROCEDURE

The proposed modeling procedure clearly describes the steps to build estimation models. The six steps in this procedure are displayed in Figure 1. Moreover, the tasks and their importance are also explained in the following sections.

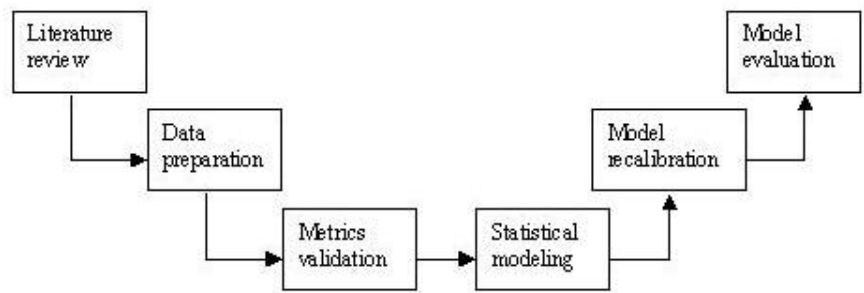

Figure 1.    A general procedure to build the model

### 3.1.  Literature Review

In this step, candidate software metrics and possible techniques are reviewed and selected for building software quality estimation models. The problem with design and complexity metrics is that those metrics can only be collected in later stages of the development cycle. For building estimation models that can be used in early stages, software metrics describing the characteristics of the project should be analyzed for quality estimation. Besides traditional statistical modeling techniques, many machine learning and soft computing techniques are also appropriate candidates for producing software quality estimation models.

### 3.2.  Data Preparation

Data need to be collected, which include software metrics and software quality information. Since quality-oriented information is usually protected by the companies and other organizations, there is little data in public domain. Possible ways to obtain this kind of data are to purchase it from some organizations or send out questionnaires and wait for responses. When the data is available, preliminary data analysis is needed to exclude redundant and unrelated information. Moreover, certain form of data transformation may be necessary to make the data more suitable for further processing.

### 3.3. Metrics Validation

Since not all factors collected were actually related to the quality metrics, they have to be validated to be included in the estimation model. Correlation analysis, ANOVA and machine learning are possible techniques to determine the metrics, which will be explained thoroughly in Sections 4.1, 4.2 and 4.3. For quantitative and qualitative variables, different techniques have to be applied. Furthermore, the results should be examined conservatively to avoid missing any predictors.

### 3.4. Statistical Modeling

Applicable regression techniques need to be explored to fit the data and establish a model. First the distribution characteristics should be analyzed to determine the form of the formula. Under most circumstances it is a tough job, and experiments have to be done to test the suitability of a certain form. Then corresponding regression techniques need to be applied to calculate the parameters in the formula. Some suitable methods are discussed in Section 4.4.

### 3.5. Model Recalibration

The regression models have some limitations, such as less generalization. Also, the imprecise information of the characteristics of the software project can be treated in more sophisticated ways. Therefore soft computing techniques can be applied to further calibrate the parameters in the models. A neuro-fuzzy approach is proposed to improve the estimation accuracy, as described in Section 4.5. The fuzzy part tries to deal with vague inputs and expert knowledge of software metrics and the neural part is to learn the parameters from data to achieve better performance.

### 3.6. Model Evaluation

The models derived from the previous steps have to be evaluated according to certain criteria to compare the performance. Cross-validation is adopted to achieve average results of the criteria such as MMRE and Pred(m). A brief introduction is presented in Section 4.6.

## 4. TECHNIQUES ADOPTED

The techniques utilized to achieve the research objective include: correlation analysis, ANOVA, machine learning, regression, soft computing and cross-validation. These techniques are applied in different steps of the procedure.

## 4.1. Correlation Analysis

Firstly, the relationships between each explanatory variable (predictor) and the dependent variable (the number of defects) are revealed by using correlation coefficients. Although correlation between two variables does not necessarily result in causal effect, it is still an effective method to choose candidate quantitative metrics. Since not all the quantitative explanatory variables follow normal distributions, Spearman's rank correlation coefficient method was chosen instead of standard correlation coefficient methods, where ranks of the variables are used rather than real values [16].

## 4.2. ANOVA

There are many related factors under research that are qualitative. For these factors, one way Analysis of Variance (ANOVA) can be performed to find out how effective they are for the dependent variable [17]. The data samples can be treated as independent groups according to the classifications of those categorical (qualitative) variables, and one-way ANOVA is the vehicle to test the differences among the groups. The differences indicate how the variables influence the dependent variable.

## 4.3. Machine Learning

To confirm the results of metrics validation from statistical methods, machine learning can provide more evidence to evaluate the significance of the factors. Machine learning refers to a system that has the capability to automatically learn knowledge from experience and other ways [18]. The problem we try to resolve is to determine the significant software metrics for quality estimation. Knowledge discovery techniques in machine learning area are suitable to accomplish the task. In this paper we use decision trees to explore the relationships between the quality factors and the quality metric. A tree structure is established by using certain tree generation algorithm of decision trees, which has independent variables as interior nodes and possible classification of dependent variable as leafs [19]. The structure thus represents the relevancy of the variables and is easy to understand and interpret.

## 4.4. Regression

Many issues have to be addressed for building the regression model. Firstly, the form of the regression model should be decided. Generally, the number of defects found in the software lifecycle is regarded as a dependent variable, which has various forms of relationships with the predictors. Although the function is not deterministic, it can be treated like a linear one with appropriate transformation [20].

Then certain regression technique should be explored to determine the parameters in the function. Although the function can take a linear form, ordinary linear regression is not applicable here because many predicators (factors) are qualitative. Usually dummy variables have to be designed to apply traditional linear regression, but the results would be hard for interpretation and impossible for further recalibration. A special approach named CATREG (Categorical regression with optimal scaling using alternating least squares) is suitable to assign numerical values to those qualitative variables and obtain the final regression formula [21]. The rationale behind it is transforming the categorical variables according to the optimal scaling levels (nominal or ordinal) and optimizing the quantifications following the least square criterion [22].

## 4.5. Soft Computing

The main strength of neural networks is their power to model extremely complex systems. Examples include the cases where there are a great number of variables and traditional regression methods are not applicable. Moreover, neural networks can learn from historical data and train themselves to achieve high performance, where not much expertise is mandatory.

Another sophisticated technique for modeling complex systems is fuzzy logic, which is applied with heuristic knowledge and with imprecise inputs to realize complicated functions. Because the real world is full of vagueness, fuzzy logic has proved to be very successful in many fields, like dynamic control, decision support, and other expert systems.

Both neural network and fuzzy logic have their advantages in modeling nonlinear functions, but their strengths attract us from quite different aspects. Moreover, since neural network lacks the capability of explaining the meaning behind the application, which is exactly the strength of fuzzy logic, it has drawn great attention to combine the two techniques together to form a new approach. Here comes neuro-fuzzy, which is now an important constituent of soft computing, and is applied to solve decision-making, modeling, predicting, and control problems in the real world.

In this paper, we adopt ANFIS (Adaptive Neuro-Fuzzy Inference System) to combine the advantages of neural networks and fuzzy logic to further tune the quantifications in the regression model [23]. ANFIS is actually a fuzzy inference system, which is based on a neural network structure, and thus it has the learning capability inherited from neural networks to adjust the parameters of membership functions and rules. A typical structure of ANFIS is shown in Figure 2.

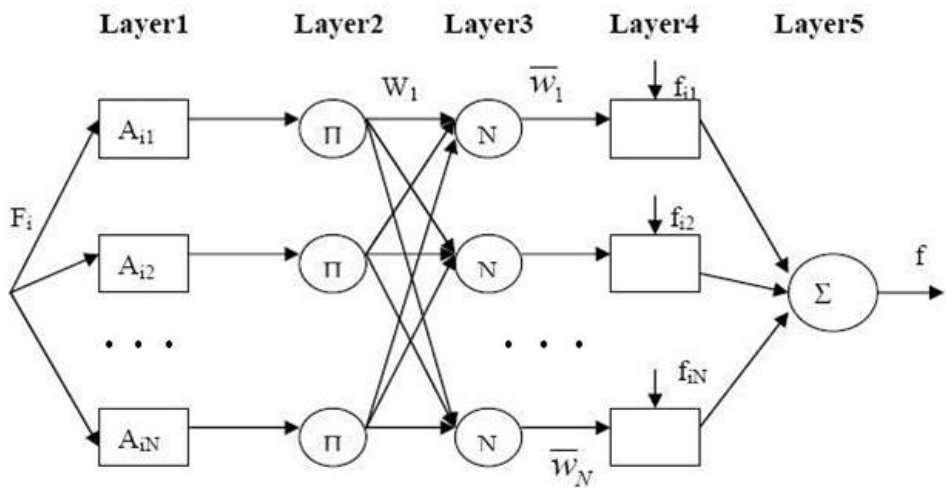

Figure 2: A typical structure of ANFIS

The purpose of embedding ANFIS in the recalibration is to fine tune the quantifications of those qualitative variables in order to achieve better estimation performance. Therefore the actual structure of the recalibration approach is designed as in Figure 3. Each neuro-fuzzy agency (NFA) is a single ANFIS for each of the qualitative variable (F$i$). Every qualitative variable goes through its respective NFA to obtain the recalibrated value (NF$i$) for the algorithmic model, and the quantitative variables (V$i$) are fed directly into the algorithm model without recalibration [24,25]. Then the estimation of the model is calculated. The training process can be treated as a back-propagation method, with the error signals being fed back into each of NFA and the parameters of the NFA then being adjusted following the Delta rules. After the training converges, the trained consequent parameters of each NFA are taken as the final quantifications of that qualitative variable. For future applications of the software quality estimation model, intermediate values among those assigned groups of the categorical variables (metrics) can be designated to gain higher estimation accuracy.

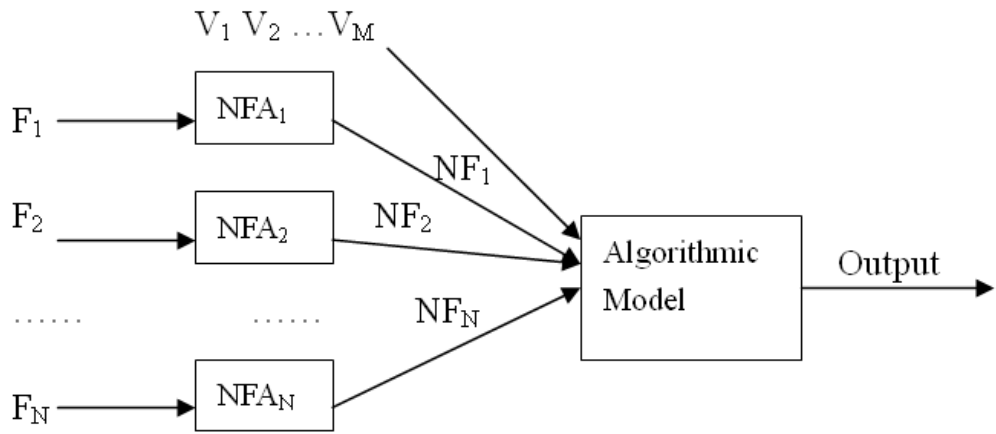

Figure 3.    Structure of the recalibration approach

### 4.6.  Cross-validation

Cross-validation is a method to make good use of limited data samples [26]. K-fold cross-validation is commonly utilized, in which the dataset is divided into K subsamples. One subsample is reserved as validation (testing) data, while the remaining K-1 subsamples are used as training data to build the model. To evaluate the performance of the built models, we have to set up the evaluation criteria first, usually MMRE and Pred(m) [27]. The cross-validation process is to be repeated K times, and the K results are then averaged to find out the final performance of that model.

## 5. EMPIRICAL RESULTS

We apply this procedure to the data from the ISBSG repository to verify the applicability of the proposed procedure.

## 5.1. Data Source

ISBSG (International Software Benchmarking Standards Group) is a non-profit organization that collects and maintains three repositories of IT historical data. Although the data it collects is not quality-oriented, we can still gather useful information from its repositories. And the repository that we are interested in is the Software Development and Enhancement Project Data. The current version is Release 10, which contains over 4,100 projects. ISBSG uses a complicated questionnaire to collect project information from companies all over the world. The questionnaire contains questions about project process, technology, people and work effort, product, function points and other related information.

## 5.2. Experimental Results

### 5.2.1. Data Preparation

Since the ISBSG repository contains data of different quality levels, to ensure that the results are convincing, we only chose data points of quality levels 'A' and 'B', which meant the contributors were seemingly confident with the data. Because Function Point is the size measurement in this research, only data points of Function Point counting levels 'A' and 'B' were selected. Moreover, the research target is software quality, and data points without defect information were excluded. There were 336 qualified data points.

There are 132 questions in the ISBSG questionnaire, under six categories, i.e. project process, technology, people and work effort, product, function points and project completion. After further examination, candidate factors/predictors were determined among more than 100 items, which were Function Points, VAF (Value Adjustment Factor), efforts, development types, experience, organization types, platforms, programming languages, maximum team size, concurrent users and etc. Since VAF is calculated from 14 GSCs (General System Characteristics), it is a good indicator to represent the variety of the projects. Unfortunately, there were only 64 data points containing valid VAF information. Thus our experiments had to base on those data points.

The major numeric variables under consideration were the number of defects and Function Points, but both of them did not follow normal distribution for linear regression. Hence natural logarithmic transformation was applied, and QQ-plot was used to validate the results.

### 5.2.2. Metric Validation

#### 5.2.2.1. Correlation Analysis

We treated some of the factors as numeric values and analyze the correlation with the number of defects. From the results of correlation analysis (Table 1), Function Points, Effort and Maximum Team Size showed close correlation with the defects at the 0.05 significant level. Therefore we decided to include these factors in the following regression modeling.

**5.2.2.2. ANOVA**

ANOVA was used to reveal the impacts that those nominal variables exert on the number of defects. Although VAF was not proved to be an effective factor as a numeric variable, it was also treated as a nominal variable since there were only several distinct values.

Table 1.    Correlation results

| Factors | Correlation coefficients | Sig. (2 tailed) |
|---|---|---|
| Function points | 0.556 | **0.000** |
| Effort | 0.640 | **0.000** |
| Maximum team size | 0.536 | **0.000** |
| VAF | 0.079 | 0.534 |

Based on the results presented in Table 2, different Organization Types, Development Platforms and Language Types could not classify the data into respective groups with different quality levels. Thus they did not present significant influences on the final defects and would not be include in the model.

Table 2.    ANOVA results

| Factors | F Value | Sig. |
|---|---|---|
| VAF | 3.979 | **.000** |
| Development Types | 12.933 | **.000** |
| Organization Types | 1.538 | .088 |
| Development Platforms | 0.157 | .925 |
| Language Types | 0.292 | .590 |

We chose Development Type and VAF for further exploration. For VAF, it was actually regarded as a numeric attribute in the regression. For Development Type, there were three nominal values, and according to the comparisons listed in Table 3, there was no significant difference between New Development and Re-development (the value of Sig. was 0.579). Therefore we combined these two categories together and treated this variable as a binary one.

Table 3.    Multiple Comparisons of Development Types

| Development Type VS. Defects (Tukey HSD) | | | |
|---|---|---|---|
| Development Type(I) | Development Type(J) | Difference (I-J) | Sig. |
| Enhancement | New Development | -267.915 | **.000** |
| | Re-development | -463.402 | **.047** |
| New Development | Re-development | -195.487 | .579 |

### 5.2.2.3. Machine Learning

We used a model tree to explore the relationships between the dependent variable (defects) and those predictors. The candidate predictors were determined based on the results from previous sections, i.e. Function Points, Efforts, VAF, Development Type, Maximum Team Size and etc. There were only two leafs (branches) in the model tree, where the splitting attribute is Efforts. For the two leafs, only Function Points, VAF, Development Type and Efforts were included in the linear formula, and the coefficients of Efforts were both near 0.

We also tried other combinations of predictors to form the model trees, and the results were consistent, i.e. Function Points, VAF and Development Type were effective predictors but the influence of Efforts needed further investigation.

### 5.2.3. Regression

The dependent variable was the natural logarithmic form of the number of defects. We first utilized ordinary linear square (OLS) to build the model. The results showed not all the predictors are statistically significant (Table 4). Efforts and Max Team Size were not significant at the 0.05 level (0.856 and 0.909). Therefore, we continued to examine the model using stepwise linear regression.

Table 4.   OLS model

| Model | | Coefficients | Std. Coefficients | Sig. |
|---|---|---|---|---|
| | | B | Beta | |
| 1 | (Constant) | -6.615 | | **.010** |
| | Function Points | .864 | .501 | **.011** |
| | VAF | 6.066 | .393 | **.010** |
| | Efforts | -.051 | -.039 | .856 |
| | Development Type | -1.300 | -.377 | **.016** |
| | Max Team Size | .004 | .017 | .909 |

The results of stepwise linear regression were listed in Table 5 (showing regression coefficients in the stepwise steps). Model 3 was the final model chosen by the stepwise linear regression.

The results indicated that only Function Points, VAF and Development Type entered as independent variables, and the corresponding coefficients were significant. With regard to Efforts, although it presented tight correlation with Defects, it was excluded from the final formula since it is also closely correlated with one of the predictors, i.e. Function Points (Table 6). Since VAF was regarded as a numerical variable in regression and there was only one binary qualitative predictor (Development Type), CATREG was not necessary for the quantifications and the regression. The final formula obtained again with OLS was:

$$Defects = -5.939 + 0.704*FP + 6.011*VAF - 1.480*Development\ Type \qquad (1)$$

Table 5.    Stepwise linear regression results

| Model | | Coefficients | Std. Coefficients | Sig. |
|---|---|---|---|---|
| | | B | Beta | |
| 3 | (Constant) | -6.659 | | **.006** |
| | Function Points | .828 | .480 | **.000** |
| | VAF | 5.933 | .384 | **.004** |
| | Development Type | -1.278 | -.370 | **.006** |

Table 6.    Correlation between Function Points and Efforts

| Factors | Correlation | Sig. |
|---|---|---|
| FP vs. Efforts | 0.624 | **.000** |

In this formula, VAF is in the range of [0.65 1.35] and Development Type is a binary attribute. Both Defects and FP take natural logarithmic form. As mentioned before, VAF is calculated from 14 General System Characteristics (GSCs), which have a rating from 0 to 5. Thus the simplest project may have a VAF value as 0.65, and most complicated as 1.35. For the simplest New Development project (VAF is 0.65), when function point counting is less than 18, the number of final defects may come out near 1. For extreme conditions, the right hand of the formula may result in a negative number. Since the variable of Defects is in natural logarithmic form, the final number of Defects is around 0.

### 5.2.4. Recalibration

In the previous linear regression formula, only VAF could be regarded as a qualitative variable, which was applicable for further neuro-fuzzy recalibration.

A simple ANFIS (Adaptive Neuro-Fuzzy Inference System) was established to adjust VAF quantifications to make the values more effective. The quantifications of VAF have already been calculated from 14 GSCs, not from CATREG. After training, we obtained another set of VAF values, which were again used in the regression formula.

### 5.2.5. Model Evaluation

First, we carried out neuro-fuzzy recalibration using all data points for training, and compared the Pred(0.25) and MMRE values before and after the recalibration. The results showed significant improvements, 14.29% and 55.86%, i.e. Pred(0.25) from 0.2187 to 0.2500 and MMRE from 1.3204 to 0.5828. Both of the Pred(0.25) values were relatively low. The reason behind them could be the samples were not sufficient and too few predictors were included in the models.

Since it is usually not a good idea to apply all data samples for training regarding the threat of overfitting, we split the data into training and checking samples and checked the performance after training. Both 8 folds and 4 folds cross-validation were covered to evaluate MMRE of the regression and recalibrated models. Pred(m) was not comparable in this case and excluded due to insufficient data points for checking.

According to the experimental results (Table 7, 8 folds), average MMRE was reduced from 1.3204 to 0.8293. Similar to the prior one, the 4 folds results (Table 8) presented average MMRE improvement from 1.3204 to 0.9243. Both of the improvements, i.e. 32.50% and 28.51%, were significant.

Table 7.    Performance of cross-validation (8 folds)

|  | Regression (MMRE) | Recalibrated (MMRE) | Improvement |
|---|---|---|---|
| Experiment 1 | 0.6168 | 0.5722 | 7.24% |
| Experiment 2 | 2.6174 | 1.4238 | 45.60% |
| Experiment 3 | 0.7957 | 0.7291 | 8.38% |
| Experiment 4 | 1.9843 | 0.8837 | 55.47% |
| Experiment 5 | 0.798 | 0.417 | 47.74% |
| Experiment 6 | 0.8812 | 0.3971 | 54.94% |
| Experiment 7 | 1.2097 | 1.1643 | 3.75% |
| Experiment 8 | 1.66 | 1.0475 | 36.90% |
| Average | 1.3204 | 0.8293 | 32.50% |

Table 8.    Performance of cross-validation (4 folds)

|  | Regression (MMRE) | Recalibrated (MMRE) | Improvement |
|---|---|---|---|
| Experiment 1 | 0.9528 | 0.6559 | 31.16% |
| Experiment 2 | 0.7032 | 0.548 | 22.07% |
| Experiment 3 | 1.5968 | 1.2216 | 23.50% |
| Experiment 4 | 2.0288 | 1.2716 | 37.32% |
| Average | 1.3204 | 0.9243 | 28.51% |

To avoid bias during data separation, we experimented with randomly choosing certain percentage of data points (60%, 70% and 80%) for training, and the reminder for checking. This process was repeated for 10 times for each percentage. The respective MMRE improvements for the three cases were 26.70%, 30.58% and 25.70% (Table 9), and all of them were significant.

## 6. CONCLUSIONS AND FUTURE WORK

In this paper we propose a general procedure to derive software quality estimation models. In the proposed procedure, the tasks of each step are explained in details and relevant techniques are suggested to accomplish the tasks.

Table 9.   Performance of random training data points

|  | Improvement (MMRE) 60% | Improvement (MMRE) 70% | Improvement (MMRE) 80% |
|---|---|---|---|
| Experiment 1 | 28.7265% | 26.9268% | 5.4948% |
| Experiment 2 | 6.1767% | 27.2407% | 3.9235% |
| Experiment 3 | 4.3487% | 28.9030% | 51.3789% |
| Experiment 4 | 21.7967% | 30.2943% | 9.1249% |
| Experiment 5 | 31.4237% | 22.3646% | 3.9546% |
| Experiment 6 | 49.8950% | 39.7557% | 48.6786% |
| Experiment 7 | 35.7416% | 46.4799% | 26.7754% |
| Experiment 8 | 20.3925% | 2.8581% | 41.9078% |
| Experiment 9 | 24.3479% | 49.8335% | 29.8117% |
| Experiment 10 | 44.1545% | 31.0979% | 35.9454% |
| Average | 26.7004% | 30.5755% | 25.6996% |

The three research questions are addressed in this paper. Firstly, several methods, such as correlation analysis, ANOVA and decision trees, are advised to determine software metrics for software quality estimation. Secondly, we suggest applying regression approaches like ordinary least square, stepwise linear regression and CATREG to build estimation models from data, and then using neuro-fuzzy approach to recalibrate the quantifications of those qualitative parameters to improve performance. Lastly, a six-step procedure is discussed to fulfill the whole modeling objective.

The main advantage of this procedure is applying neuro-fuzzy approach along with traditional statistical modeling. Following this procedure, one can obtain a statistical model from data first, and then improve its performance with the recalibration method. On the other hand, if an algorithmic model already exists, it can be recalibrated to achieve better estimation accuracy. This procedure is also applicable to build estimation models in other areas of software engineering.

Based on data from the ISBSG repository, we verified the proposed procedure step by step. Even though the data repository is not truly perfect for this research and we could not extract enough predictors to build an effective estimation model, the whole procedure has been proved to be appropriate for deriving software quality estimation model. Most importantly, the recalibration step improved the performance of the previously derived estimation model. The empirical value of this procedure can also be extended to other estimation problems.

Possible directions of our future work include the following three aspects:

(1) If detailed GSC information can be obtained for the projects, a quality estimation model can be established based on these predictors instead of VAF. Hence, more effective quality metrics can be involved for more accurate estimation;

(2) Since it is difficult to get industrial data on software quality, information of open source software (OSS) projects may be a good alternative. We plan to collect OSS data to build software quality estimation model and compare the different practices.

(3) The distribution of defects throughout software development lifecycle is also a key issue in software quality assurance. We plan to model the spread of defects in different phases based on the total number of defects and the characteristics of the project.

## ACKNOWLEDGEMENTS

The authors would like to thank ISBSG for providing the data repository to accomplish the research.

**Authors:**

Jie Xu is currently a Ph.D. student with the Electrical and Computer Engineering Department, the University of Western Ontario. His research interests include software engineering, software estimation, soft computing, and data mining. His email address is jxu89@uwo.ca.

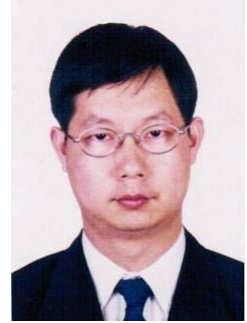

Danny Ho is an independent management consultant and advisor to two startup companies. He is also appointed as an Adjunct Research Professor at the Department of Software Engineering, Faculty of Engineering, The University of Western Ontario and University of Ontario Institute of Technology. His areas of special interest include software estimation, project management, object-oriented software development, and complexity analysis. He is currently a member of the Professional Engineers Ontario (PEO) and a Project Management Professional (PMP). His email address is danny@nfa-estimation.com.

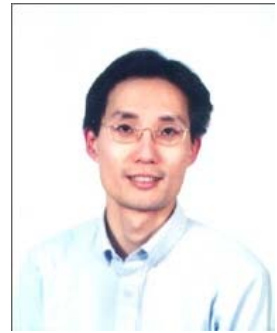

Luiz Fernando Capretz is currently an Associate Professor and the Director of the Software Engineering Program at the University of Western Ontario, Canada. His present research interests include software engineering (SE), human factors in SE, software estimation, software product lines, and software engineering education. He is an IEEE senior member, ACM distinguished member, MBTI certified practitioner, Professional Engineer in Ontario (Canada). He can be reached at lcapretz@eng.uwo.ca.

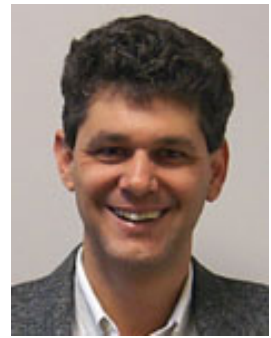